\documentclass[prd,twocolumn,showpacs,preprintnumbers,amsmath,amssymb]{revtex4}
\usepackage{graphicx}
\begin{document}
%\preprint{DRAFT}

%\setlength{\topmargin}{-0.25in}
\title{Series expansions of the density of states in $SU(2)$ lattice gauge theory}
\author{A. Denbleyker}
\email[]{alan-denbleyker@uiowa.edu}
\author{Daping Du}
\email[]{daping-du@uiowa.edu}
\author{Yuzhi Liu}
\email[]{yuzhi-liu@uiowa.edu}
\author{Y. Meurice}
\email[]{yannick-meurice@uiowa.edu}
\affiliation{Department of Physics and Astronomy\\ The University of Iowa\\
Iowa City, Iowa 52242, USA }
\author{A. Velytsky}
\email[]{ vel@theory.uchicago.edu}
\affiliation{Enrico Fermi Institute, University of Chicago, 5640 S. Ellis Ave., Chicago, IL 60637, USA,}
\affiliation{HEP Division and Physics Division, Argonne National Laboratory, 9700 Cass Ave., Argonne, IL 60439, USA}

\date{\today}
\begin{abstract}
We calculate numerically the density of states $n(S)$ for $SU(2)$ lattice gauge theory on $L^4$ lattices. Small volume dependence are resolved for small values of $S$. 
We compare $ln(n(S))$ with weak and strong coupling expansions. Intermediate order 
expansions show a good overlap for values of $S$ corresponding to the crossover.
We relate the convergence of these expansions to those of the average plaquette.
We show that when known logarithmic singularities are subtracted from $ln(n(S))$, 
expansions in Legendre polynomials appear to converge and could be suitable to determine the Fisher's zeros of the partition function.
\end{abstract}
\maketitle

\section{Introduction}

Quantum Chromodynamics is a widely accepted theory of strong interactions. 
From a theoretical point of view, understanding the large distance behavior in terms 
of the weakly coupled short distance theory has been an important challenge. 
The connection between the two regimes can be addressed meaningfully using the lattice formulation. In the pure gauge theory (no quarks) described with the standard Wilson's action, no phase transition between the weak and strong coupling regime has been found numerically for $SU(2)$ or $SU(3)$ and the theory should be in the 
confining phase for all values of the coupling. Recently, convincing arguments have 
been given \cite{tomboulis07,tomboulis07b} in favor of the smoothness of the renormalization group flows between the two fixed points of interest, putting the confining 
picture on more solid mathematical ground. 

\begin{figure}
\includegraphics[width=\columnwidth]{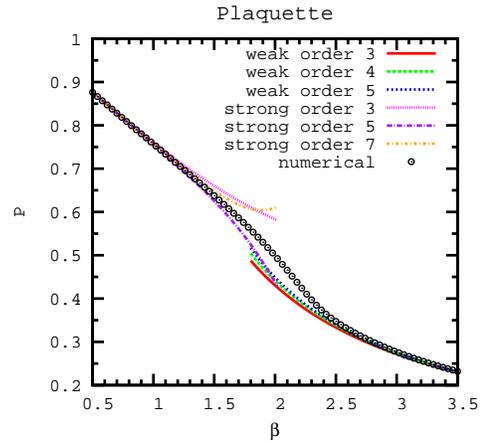}
\caption{
\label{fig:pall}  Weak and strong coupling expansions of the average plaquette $P$ 
for $SU(2)$ at various orders in the weak and strong coupling expansion compared to the numerical values. }
\end{figure}

The absence of phase transition discussed above suggests that it is possible to match 
the weak coupling and the strong coupling expansions of the lattice formulation. 
However, if we consider these two expansions, for instance for the average $SU(2)$ plaquette as a function of $\beta=4/g^2$, we see in Fig. \ref{fig:pall} that there is a 
crossover region 
(approximately $1.5<\beta<2.5$) where none of the two expansions seem to work.
This behavior is probably related to singularities in the complex $\beta$ plane \cite{kogut80b,third} 
that are not completely understood. In the case of the one plaquette model \cite{plaquette}, taking the inverse Laplace transform with respect to $\beta$ (Borel transform) of the partition function yields a function that has better convergence properties.  It would be interesting to know if this feature persists on $V=L^4$ lattices.

In this article, 
we study expansions of the inverse Laplace transform of the partition function 
(the density of states) of $SU(2)$ lattice gauge theory on symmetric 4 dimensional lattices. The density of states is denoted $n(S)$ and  defined precisely in Sec. \ref{sec:density}. It gives a relative measure of the number of ways to get a value $S$ 
of the action. Knowing $n(S)$, we can calculate the partition function and its derivatives for any real or complex value of $\beta$.  In particular, it could be used to determine the Fisher's 
zeros of the partition function \cite{alves90b,quasig,lat07}.
The choice of $SU(2)$ is motivated by the existence of a particular 
symmetry \cite{gluodyn04} which allows to determine the behavior of $n(S)$ near 
its maximal argument without extra calculation. 
In Sec. \ref{sec:vol}, we explain  why $ln(n(S))$ is expected to scale like the volume and can be interpreted as a "color entropy". Numerical calculations of $n(S)$ obtained by patching plaquette distributions 
multiplied by the inverse Boltzmann weight at 
values of $\beta$ increasing by a small increment are presented in Sec. \ref{sec:num}.
The article is focused on comparisons with numerical data on a $6^4$ lattice  where 
finite volume effects are not too large and plaquette distributions broad enough 
to allow a smooth patching. The values of $n(S)$ on such lattice are compared 
with those on a $4^4$ and $8^4$ lattice. It is interesting to note that the volume dependence is resolvable only for small values of $S$ where a behavior $ln(S)/V$ 
is observed for $ln(n(S))$ 

The numerical results are compared with expansions that can be obtained from 
the strong (Sec. \ref{sec:strong}) and weak (Sec. \ref{sec:weak}) coupling 
expansions of the average plaquette. Intermediate orders in these expansions show 
a good overlap for values of $S$ that correspond to the crossover.  We then show that 
the convergence of the new series can be related empirically to those of the series 
for the average plaquette.  The weak coupling expansion determines the logarithmic 
singularities of $ln(n(S))$ at both boundaries. When these singularities are subtracted 
we obtain a bell-shaped function that can be approximated very well by Legendre polynomials (Sec. \ref{sec:leg}). We conclude with possible applications for the 
calculations of the Fisher's zeros and open problems.
\section{The density of states}
\label{sec:density}
We  consider the standard pure gauge partition function
\begin{equation}
Z=\prod_{l}\int dU_l {\rm e}^{-\beta \mathcal{S}} \  ,
\end{equation}
with the Wilson action
\begin{equation}
\mathcal{S}=\sum_{p}(1-(1/N)Re Tr(U_p))\  .
\end{equation} 
and $\beta\equiv2N/g^2$. 
We use a $D$ dimensional cubic lattice with periodic boundary conditions. 
For a symmetric lattice with $L^D$ sites, the 
number of plaquettes is 
\def\mn{\mathcal{N}_p}
\begin{equation}
\mathcal{N}_p\equiv\ L^D D(D-1)/2\  .
\end{equation} 
In the following, we restrict the discussion to the group $SU(2)$ and $D=4$.  
%We will also assume that the lattice has an even number of sites in each direction.
For $SU(2)$, one can show \cite{gluodyn04} that the maximal value of $\mathcal{S}$ is  $2\mn$. 
We define the average plaquette: 
\begin{equation}
\label{eq:pdef}
P\equiv\left\langle\mathcal{ S}/\mn\right\rangle =-d(lnZ/\mn)/d\beta \ .
\end{equation}

Inserting $1$ as the integral of 
delta function over the numerical values $S$ of $\mathcal{S}$  in $Z$, we can write
\begin{equation}
Z =\int_0^{2\mn}dSn(S){\rm e}^{-\beta S}\ ,
\label{eq:intds}
 \end{equation}
 with
 \begin{equation}
n(S)=\prod_{l}\int dU_l \delta(S-\sum_{p}(1-(1/N)Re Tr(U_p)))
\end{equation}
 We call $n(S)$ the density of states . 
 A more general discussion 
 for spin models \cite{alves89} or gauge theories \cite{alves91} can be found in the literature where the density of states is sometimes called the spectral density.  
 From its definition, it is clear that $n(S)$ is positive. 
 Assuming that the Haar measure for the links 
 is normalized to 1, the partition function at $\beta=0$ is 1 and consequently we 
 can normalize $n(S)$ as  a probability density. 
 
 A first idea regarding the convergence properties of various expansions 
 can be obtained from the single plaquette 
 model \cite{plaquette}. In that case, we have 
 \begin{equation}
 n_{1 pl.}(S)=\frac{2}{\pi}\sqrt{S(2-S)}\  .
 \end{equation} 
The large $\beta$ behavior of the partition function is determined by the behavior of 
$n(S)$ near $S=0$. In this example, $n(S)\propto \sqrt{S}$ for small $S$, implies 
that $Z\propto \beta^{-3/2}$ at leading order. Successive subleading corrections can be 
calculated by expanding the remaining factor $\sqrt{2-S}$ in powers of $S$ and integrating over $S$ from 0 to $\infty$. 
If we factor out the leading behavior, we obtain a power series  in $1/\beta$.
The large order behavior of this power series is determined by the large order behavior of the expansion of  $\sqrt{2-S}$, itself dictated  
by the branch cut at $S=2$. One can see \cite{plaquette} that the $S$-integration over the {\it whole} positive real axis converts an expansion with a finite radius of convergence into one with a zero radius of convergence.  On the other hand, if the $S$-integration is 
carried over the interval $[0,2]$, the resulting series converges but the coefficients 
need to be expressed in terms of the incomplete gamma function. 
From this example, one may believe that it is easier to approximate $n(S)$ than the 
corresponding partition function.
However, it is not clear that these considerations will survive the infinite volume limit. 
Note also that the behavior of $n(S)$ near $S=2$ can be probed by taking $\beta 
\rightarrow -\infty$ in agreement with the common wisdom that the large order behavior 
of weak coupling series can be understood in terms of the behavior at small negative 
coupling. 

It was showed \cite{gluodyn04} that if the lattice has even number of sites in each direction and if the gauge group contains $-\openone$,  that it is possible to change $\beta Re TrU_p$ into $-\beta Re TrU_p$ by a  change of variables $U_l\rightarrow -U_l$ 
on a set of links such that for any  plaquette,  exactly one link of the set belongs to that plaquette.  This implies
\begin{equation}
	Z(-\beta)={\rm e}^{2\beta\mathcal{N}_p}Z(\beta)
	\label{eq:su2id}
\end{equation}
This symmetry implies that 
\begin{equation}
n(2\mn -S)=n(S)\  .
\label{eq:dual}
\end{equation}
In the following, we will be working exclusively with $SU(2)$ which contains  $-\openone$
and lattices with even numbers of sites in every direction. We will thus assume that 
Eq. (\ref{eq:dual}) is satisfied and we only need to know $n(S)$ for $0\leq S\leq \mn$.

\section{Volume dependence}
\label{sec:vol}
In this section, we discuss the volume dependence of the density of state. We 
make this dependence explicit by writing $n(S,\mn)$.  Given the density of states, we 
can always write 
\begin{equation}
f(x,\mn)\equiv ln( n(x\mn,\mn))/\mn\  .
\end{equation}
The function is nonzero only if $0\leq x \leq 2$. 
The symmetry (\ref{eq:dual}) implies that 
\begin{equation}
f(x,\mn)=f(2-x,\mn)
\end{equation}

In the statistical mechanics interpretation of the partition function (where $\beta$ is 
an inverse temperature), $f(x,\mn)$ can be interpreted as a density of entropy.
The existence of the infinite volume limit requires that 
\begin{equation}
lim_{\mn \rightarrow \infty} f(x,\mn) = f(x)\ ,
\end{equation}
with $f(x)$ volume independent. In the same limit, the integral ( \ref{eq:intds}) can be 
evaluated by the saddle point method. The maximization of the integrand requires 
\begin{equation}
f'(x)=\beta \  .
\label{eq:saddle}
\end{equation}
We believe that $f(x)$ is strictly  increasing for $0<x<1$ with an absolute maximum 
at $x=1$. By symmetry, this would imply that  $f(x)$ is strictly  decreasing for $1<x<2$. 
We also believe that $f'(x)$ is strictly decreasing and that Eq. (\ref{eq:saddle}) has 
a unique solution (with positive $\beta$ if $0<x<1$ and negative $\beta$ if $1<x<2$).
The numerical study of Sec. \ref{sec:num} is in agreement with these statements, 
but we are not aware of mathematical proofs.
Assuming that Eq. (\ref{eq:saddle}) has 
a unique solution, the infinite volume solution should be $x=P$ the average plaquette defined above.
We can then convert an expansion for $P$ into an expansion of $f$. 
If we want to include the volume dependence, the distribution has a finite width,
and we should expand about the saddle point and perform the integration. 
In the following, we will work at large but finite volume and residual volume dependence 
in $f$ will be kept implicit in equations.

The behavior of $f(x)$ for small $x$, can be probed by studying the model at large 
positive $\beta$ (weak coupling expansion discussed in Sec. \ref{sec:weak}).
On the other hand, at small values of $\beta$ (strong coupling expansion discussed in Sec. \ref{sec:strong}), the partition function is dominated 
by the behavior of $f(x)$ near its peak value $x=1$.
For convenience, we introduce notations suitable for the study of the density of state 
near $x=1$
\begin{equation}
g(y)\equiv f(1+y)\  .
\label{eq:gdef}
\end{equation}
$g(y)$ is then an even function defined for $-1<y<1$.  

\section{Numerical Calculation of $n(S)$}
	\label{sec:num}
\begin{figure}
		\includegraphics[width=\columnwidth]{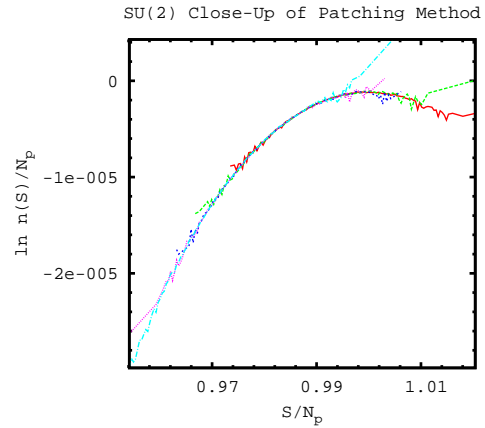}
			\caption{
				\label{fig:patch_zoom}Close-up of the patching process for $6^4$.
			}
	\end{figure}

\begin{figure}
		\includegraphics[width=\columnwidth]{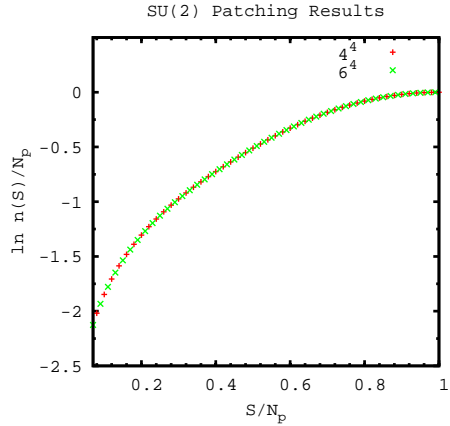}
			\caption{
				\label{fig:patching}Results of patching for $4^4$ and $6^4$.
			}
	\end{figure}

To find $n(S)$ numerically we will use a Monte Carlo simulation to create configurations of SU(2) for different values of $\beta$.  In the following example we will follow the steps we will use to find $n(S)$ for a volume of $6^4$.  We will start with 550 different sets of data ranging from $\beta = 0.02$ to $\beta = 11.00$ in steps of 0.02 and with sizes of $10^5$ configurations.  To join the data from different 
values of $\beta$ we will first create histograms of each set of data, each of these histograms is roughly Gaussian in shape.  We then filter out the data that has statistics that are lower than half of the maximum bin.  We can then remove the beta dependence by multiplying the height of each bin by ${\rm e}^{\beta S}$.  We will be left with a series of arches which when overlayed on each other form the curve $n(s)$.  To create this overlay we will start with the lowest $\beta$, which will correspond to the peak of $n(s)$, and then take the logarithm of this.  We will then look at the neighboring $\beta$ and do the same thing but then shifting it up or down so that the average distance in the bins overlapping with the first is zero.  We will then continue in this manner until the supply of datasets has been exhausted.  A portion of this process can be seen in Fig. \ref{fig:patch_zoom}.  We then average the points for each bin together and divide both the bin width and height by $\mathcal{N}_{p}$ and shift the top of the curve to zero to make the final output, which can be seen in Fig. \ref{fig:patching} for both $4^4$ and $6^4$.
We see that they overlap well.  

\begin{figure}
		\includegraphics[width=\columnwidth]{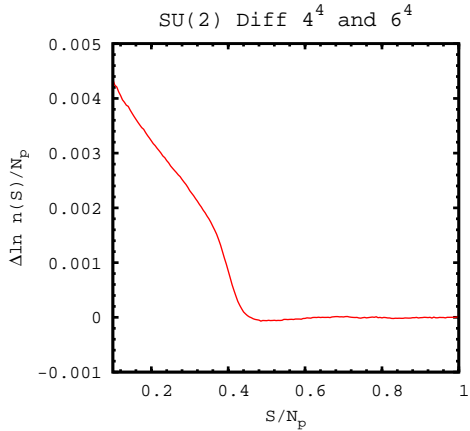}
			\caption{
				\label{fig:padiff_full}The difference between ln$(n(S))/\mn$ 
					for $4^4$ and $6^4$.
			}
	\end{figure}

\begin{figure}
		\includegraphics[width=\columnwidth]{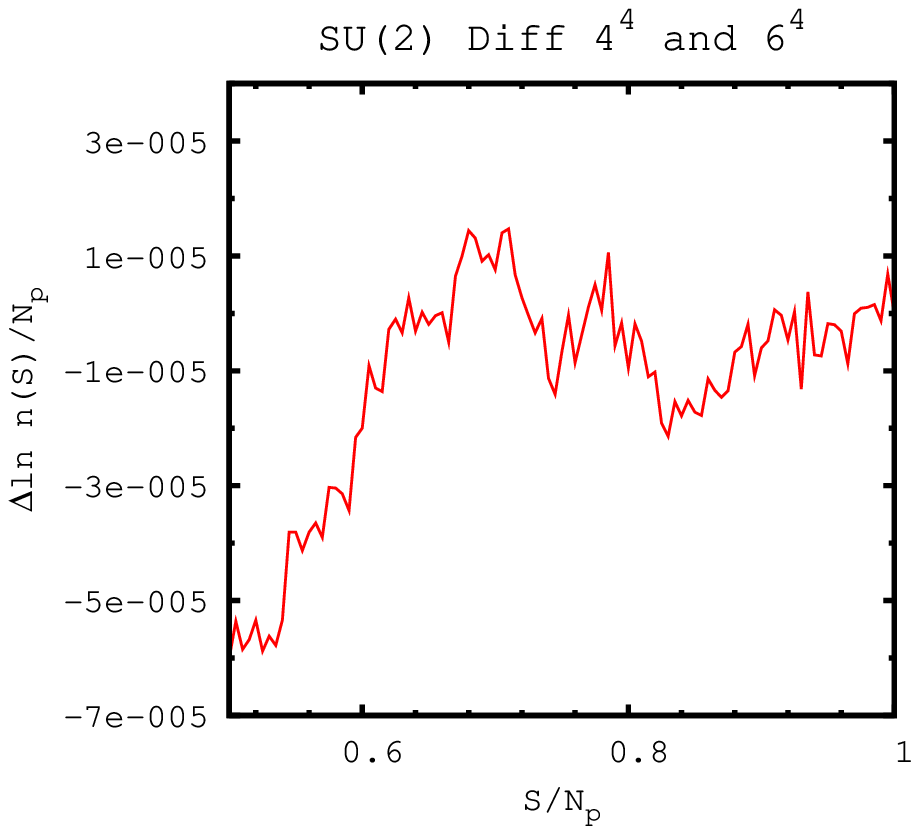}
			\caption{
				\label{fig:padiff_gt0.5}The difference between ln$(n(S))/\mn$ 
					for $4^4$ and $6^4$ with $\beta>0.5$ (close up).
			}
	\end{figure}

We now consider the difference between two different volumes, as shown in Fig. \ref{fig:padiff_full}.  We can see in Fig. \ref{fig:padiff_gt0.5} that as we get closer to $S/\mathcal{N}_p = 1$  this difference turns into noise, and as we get closer to $S/\mathcal{N}_p = 0$ we see a volume dependence growing. 
The results reported here correspond to the difference between $6^4$ and $4^4$. 
We have also studied the difference between $8^4$ and $6^4$ and found consistent results. Calculation at larger volumes are much more computationally expensive and require many more sets of data because of the narrow width of the distributions.
	\begin{figure}
		\includegraphics[width=\columnwidth]{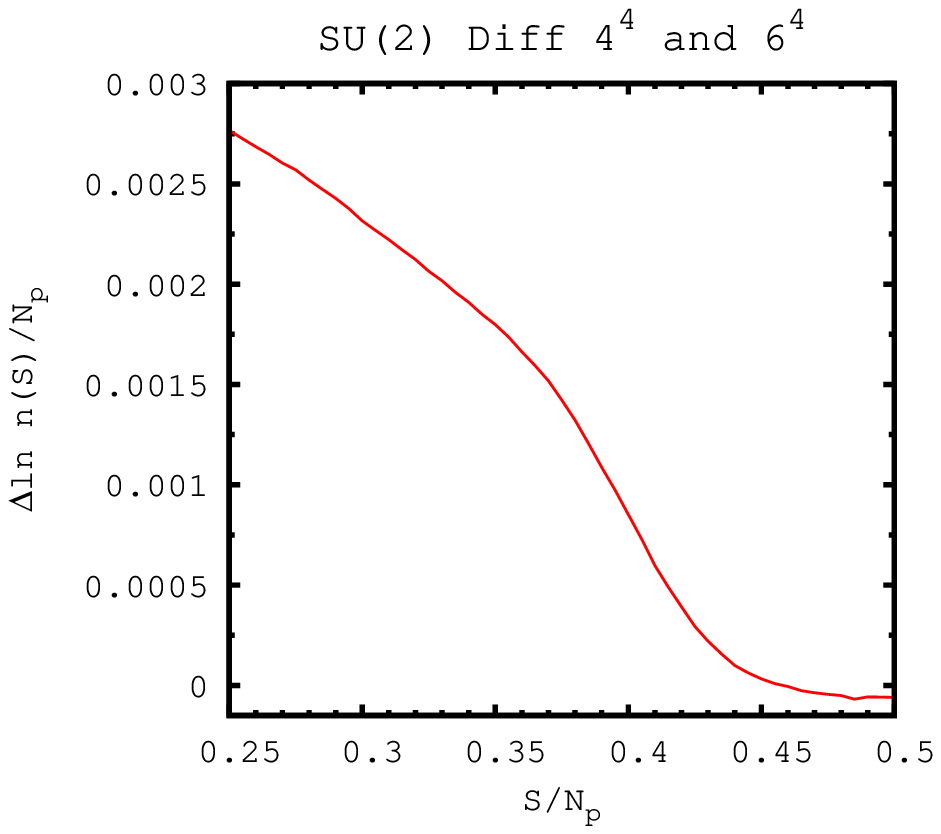}
			\caption{
				\label{fig:padiff_lt0.5}The difference between ln$(n(S))/\mn$ 
				 for $4^4$ and $6^4$ with $\beta<0.5$  (close-up).
			}
	\end{figure}
	\begin{figure}
		\includegraphics[width=\columnwidth]{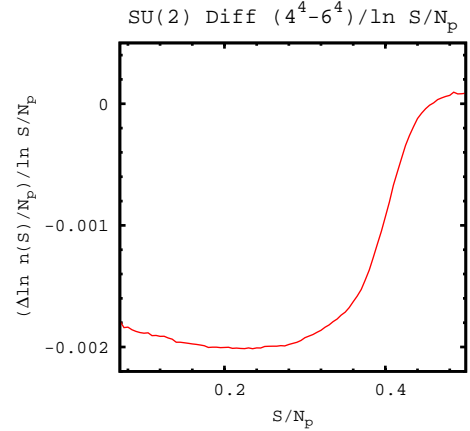}
			\caption{
				\label{fig:padiff_over_log}The difference between  ln$(n(S))/\mn$ 
			for $4^4$ and $6^4$ divided by $ln (S/\mathcal{N}_{p})$
			}
	\end{figure}

\section{Strong coupling expansion}
\label{sec:strong}
In this section, we discuss the strong coupling expansion of  
the logarithm of the density of state. We will work with the shifted function 
$g(y)$ defined in Eq. (\ref{eq:gdef}).  The strong coupling expansion of $P$ can 
be extracted from the expansion of $lnZ$ given in Ref. \cite{balian74,balian74err} 
using appropriate rescalings (for instance the $\beta$ used there is one half of the $\beta$ used here). The expansion is of the form 
\begin{equation}
P(\beta)\simeq1+ \sum_{m=1}a_{2m-1} \beta^{2m-1}  \  .
\end{equation}
The values of the coefficients are given in Table \ref{tab:sc}. 

With periodic boundary conditions, the low order coefficients are volume 
independent. This can be understood from the exact translation invariance for the 
low order strong coupling graphs that provides a multiplicity that cancels exactly the $1/\mn$ in Eq. (\ref{eq:pdef}).
Volume dependence may appear for graphs wrapping around the torus. The simplest such graph is a straight line that closes into itself due to the periodic boundary condition. 
It appears at order $\beta ^{2L}$ and has a reduced translation multiplicity since 
translation along the graph does not generate a new graph. This type of graphs 
produce $1/L$ corrections that to the best of our knowledge have not been studied 
quantitatively.  In the following, we will ignore such effects, but a study of the 
contribution of graphs with a nontrivial topology would certainly be interesting.

We will plug the expansion of $P$ in the expansion 
\begin{equation}
g(y)\simeq \sum_{m=0}g_{2m} y^{2m}\  .
\end{equation}
At lowest order we have $y\simeq a_1\beta$ and the saddle point Eq. (\ref{eq:saddle}) 
yields $2g_2 y\simeq 2g_2a_1\beta \simeq \beta$ which implies $g_2=1/(2a_1)$. 
This procedure can be followed order by order in $\beta$. 
The results are shown in Table \ref{tab:sc}.

\begin{figure}
\includegraphics[width=\columnwidth]{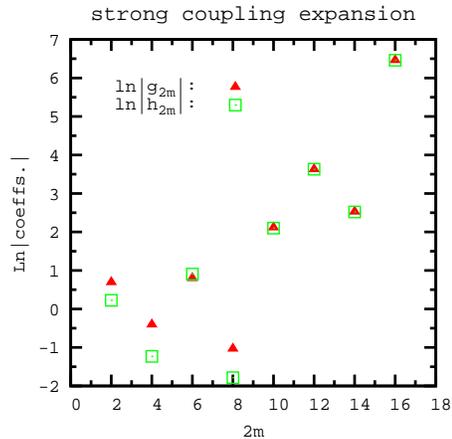}
\caption{
\label{fig:lncoefstrong} Logarithm of the absolute value of  $g_{2m}$ and $h_{2m}$}
\end{figure}

Since $n(S)$ is zero for $S=0$ and $2\mn$, we expect logarithmic singularities at 
$x=0$ and 2 for $f(x)$ and $y=\pm1$ for $g(y)$. This singularities will cause the 
strong coupling series to diverge when $|y|\geq 1$. Consequently, we define the subtracted function
\begin{equation}
\label{eq:hdef}
h(y)\equiv g(y)-A(ln(1-y^2))\ .
\end{equation}
The coefficient $A$ will be calculated using the  weak coupling expansion in Sec. 
\ref{sec:weak}. In the infinite volume limit, we have $A=3/4$.  Expanding 
\begin{equation}
h(y)\simeq \sum_{m=0}h_{2m} y^{2m}\  ,
\end{equation}
we obtained coefficients that are showed in Table \ref{tab:sc} for $A=3/4$. 
The coefficients $g_{2m}$ and $h_{2m}$ are also shown on a logarithmic scale 
in Fig. \ref{fig:lncoefstrong}.  This graph shows that the two types of coefficients
become rapidly of the same order, which indicates singularities in the complex $y$ plane for smaller values of 
$|y|$ than the ones at $\pm1$. 
\begin{table}[t]

\newcommand\T{\rule{0pt}{2.9ex}}
\newcommand\B{\rule[-2ex]{0pt}{0pt}}

\begin{tabular}{||c|c|c|c||}
\hline
 $m$\T\B&$a_{2m-1}$&$g_{2m}$&$h_{2m}$\\
 \hline
 
 1 \T &$-\frac{1}{4}$& $-2$&$-\frac{5}{4}$\\
 2 \T &$\frac{1}{96}$& $-\frac{2}{3}$& $-\frac{7}{24} $\\
 3 \T&$-\frac{7}{1536}$&$\frac{20}{9}$&$\frac{89}{36}$\\
  4\T&$\frac{31}{23040}$&$-\frac{16}{45}$&$-\frac{121}{720}$\\
  5 \T&$-\frac{4451}{8847360}$&$-\frac{16816}{2025}$&$-\frac{66049}{8100}$\\
    6 \T&$\frac{264883}{1486356480}$&$\frac{319736}{8505}$&$\frac{2566393}{68040}$\\
	 7\T&$-\frac{403651}{5945425920}$&$-\frac{3724816}{297675}$&$-\frac{14771689}{1190700}$\\
	  8\T\B&$\frac{1826017873}{68491306598400}$&$-\frac{163150033}{255150}$&$-\frac{2610017803}{4082400}$\\
  
\hline
\end{tabular}
\caption{\label{tab:sc}Strong coupling expansion coefficients defined in the text.}
\end{table}

\begin{figure}
\includegraphics[width=\columnwidth]{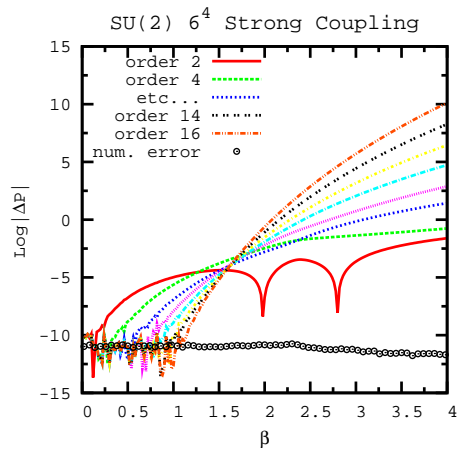}
\caption{
\label{fig:perror} Logarithm of the absolute value of the difference between the numerical data and the strong coupling expansion of $P$ at successive orders. For reference, we give the estimated numerical error on P. }
\end{figure}

\begin{figure}
\includegraphics[width=\columnwidth]{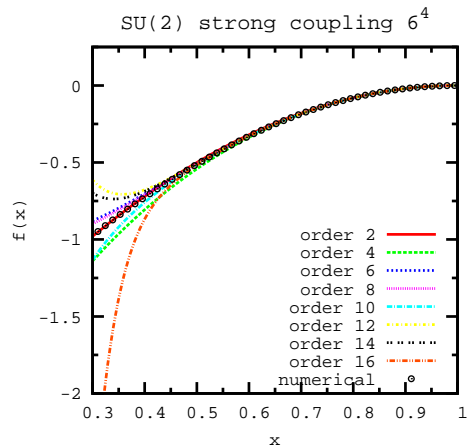}
\caption{
\label{fig:dsstrong} Numerical value of $f(x)$ compared to the strong coupling expansion at 
successive orders.}
\end{figure}

\begin{figure}
\includegraphics[width=\columnwidth]{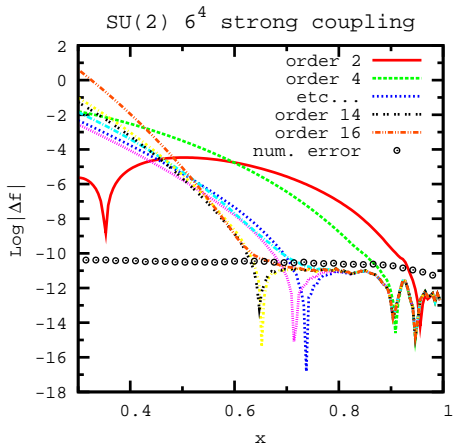}
\caption{
\label{fig:derror6strong}Logarithm of the absolute value of the difference between the numerical data and the strong coupling expansion of $f$ at successive orders. For reference, we give the estimated numerical error on $f$.}
\end{figure}

In Fig. \ref{fig:perror}, we show the error made at successive order of the strong coupling expansion  of the plaquette. We then show successive approximation of $f(x)$ 
(Fig. \ref{fig:dsstrong}) and the corresponding errors (Fig. \ref{fig:derror6strong}).

We can now compare the apparent convergence of $P$ and $f$. From Fig. \ref{fig:perror} we see that the larger order errors cross between $\beta=1.5$ and 2. For values 
of $\beta$ larger, incresing the order increases the error. This is the sign of 
a finite radius of convergence \cite{anlqcd}. Similarly, the larger order errors for $f$ cross for $x$ between 0.5 and 0.6 which are approximately the values of $P$ in the 
$\beta$ interval of crossing. Consequently, it seems like the convergence properties 
of the two expansions are the same (finite radius of convergence).

\section{Weak coupling expansion}
\label{sec:weak}
In this section, we discuss the weak coupling expansion of $f(x)$. 
The starting point is the expansion of $P$ in inverse powers of $\beta$
\begin{equation}
P(\beta)\simeq \sum_{m=1}b_m \beta^{-m}  \  .
\end{equation}
We then assume the behavior
\begin{equation}
f(x)\simeq A\ ln(x)+\sum_{m=0}f_m x^m\  .
\end{equation}
Using the saddle point  Eq. (\ref{eq:saddle}), and using the leading large $\beta$ 
and small $x$ terms we find 
\begin{equation}
\beta \simeq A/x \simeq A/(b_1/\beta)\  ,
\end{equation}
which implies $A=b_1$ at infinite volume. The procedure can be pursued order by order without 
difficulty.  The result for the 
two lowest orders is 
\begin{eqnarray}
\nonumber
 f_1 &=  & b_2/b_1  \\ \nonumber
 f_2&=   &(b_3b_1-b_2^2)/(2b_1^2)   \\ \nonumber
\end{eqnarray}
Numerical experiments indicate that the two series have the same type of growth (power 
or factorial). Note that $f_0$ cannot be fixed by the saddle point equation. The overall 
height of $f$ depends on the behavior near $x=1$ (if we insist on normalizing $n(S)$ as 
probability density) and it seems unlikely that it can be found by a weak coupling 
expansion.

At finite volume, the saddle point calculation of $P$ should be corrected in order 
to include $1/V$ effects ($V$ the number of sites , $L^D$ for a symmetric lattice). If we perform the Gaussian integration of the quadratic fluctuations, 
and use the $V$ dependent value of $b_1$ given in Eq. (\ref{eq:b1}) below, we find after a short calculation that the coefficient $A$ of $ln(x)$ is 
\begin{equation}
A=(3/4)-(5/12)(1/V)\  .
\end{equation}
This leading coefficient correction, predicts a difference of $-0.0013 ln(x)$ for the 
difference between $f(x)$ for a $4^4$ and $6^4$ and is roughly consistent 
with Fig. \ref{fig:padiff_over_log}.

Our next task is to find the values of $b_m$. 
A closed form expression can be found \cite{heller84,coste85} for $b_1$.
For the case $N_c=2$ and $D=4$, we obtain
\begin{equation}
\label{eq:b1}
b_1=(3/4)(1-1/(3V))\  .
\end{equation}
The $(-1/(3V))$ comes from the absence of zero mode $(-1/V)$ in a sum calculated in \cite{heller84} plus the contribution of the zero mode with periodic boundary conditions 
($+2/(3V)$) calculated in  
\cite{coste85}.
Numerical values for $b_2$ can be found in Ref.  \cite{heller84}
and for $b_3$ in Ref. \cite{alles93} .  In these Refs., several sums 
are calculated numerically at particular volumes that do not include $6^4$. 
Rough extrapolations from the existng data indicate that for $V=6^4$  
uncertainties are less than 0.0002 for $b_2$ and 0.0008 for $b_3$.  
For $\beta\geq 3$, these effects are close to the numerical errors for $P$. 
In the following, we use the approximate values $b_2=0.1511$ and $b_3=0.1427$ for 
$V=6^4$.

We are not aware of any calculation of $b_m$ for $m\geq 4$ for $SU(2)$. 
In the case of $SU(3)$, 
calculations up to order 10 \cite{direnzo2000} and 16 \cite{rakow05} are available 
and show remarkable regularities. 
Using the assumption \cite{third} that $\partial P/\partial \beta $ has a logarithmic singularity in the 
complex $\beta$ plane and integrating, we obtained \cite{npp} the approximate form
\begin{equation}
\sum _{m=1} b_m\beta^{-k}\approx C({\rm Li}_2 (\beta^{-1}/(\beta_m^{-1}+i\Gamma))+{\rm h.c}\ ,
\label{eq:liser}
\end{equation}
with
\begin{equation}
{\rm Li}_2(x)=\sum _{k=1}x^k/k^2 \ .
\end{equation}
We believe that at zero temperature, the new parameter $\Gamma$ which measures 
the (small) distance from the singularity to the real axis in the $1/\beta$ plane stabilizes at a nonzero value in the infinite volume limit. 
For reasons not fully understood, this parametrization of the series turns out to work very well for $SU(3)$. For instance, by fixing the value of $\Gamma $ in the middle of the allowed range 
and using the values of $b_9$ and $b_{10}$, we obtain values of the lower order 
coefficients with a relative error of 0.2 percent for $b_8$ and that increases up to 5 percent for $b_3$. In the limit $\Gamma=0$, the parametrization provides simple predictions 
for instance $b_3/b_2\simeq (4/9)\beta_m $. 
The location of the Fisher's zeros for $SU(2)$ \cite{lat07} suggests $\beta_m=2.18$.
This implies $b_3/b_2\simeq0.969$  in good agreement with our numerical estimate $b_3/b_2\simeq0.944$.  In the following we use the values $\beta_m=2.18$, 
$\Gamma=0.18/\beta _m^2\simeq 0.038$ (see \cite{lat07})  and we fixed $C=0.0062$ in order to reproduce $b_3$. The numerical values of $b_m$ and the corresponding values of $f_m$ are displayed in Table \ref{tab:wc}. 
\begin{table}[b]
\begin{tabular}{||c|c|c||}
\hline
 $m$&$b_m$&$f_m$\cr
 \hline
1&  0.7498&  0.2015\cr
2&  0.1511&  0.0999\cr
3&  0.1427&  0.0796\cr
4&  0.1747&  0.0791\cr
5&  0.2435&  0.0908\cr
6&  0.368&  0.1156\cr
7&  0.5884&  0.1597\cr
8&  0.98&  0.2351\cr
9&  1.6839&  0.3643\cr
10&  2.9652&  0.5883\cr
11&  5.326&  0.9828\cr
12&  9.7234&  1.6883\cr
13&  17.995&  2.9683\cr
14&  33.690&  5.3207\cr
15&  63.702&  9.6945\cr
\hline
\end{tabular}
\caption{\label{tab:wc} Weak coupling coefficients defined in Sec. \ref{sec:weak}. The choice of $b_1$ corresponds to $V=6^4$ .}
\end{table}

\begin{figure}
\includegraphics[width=\columnwidth]{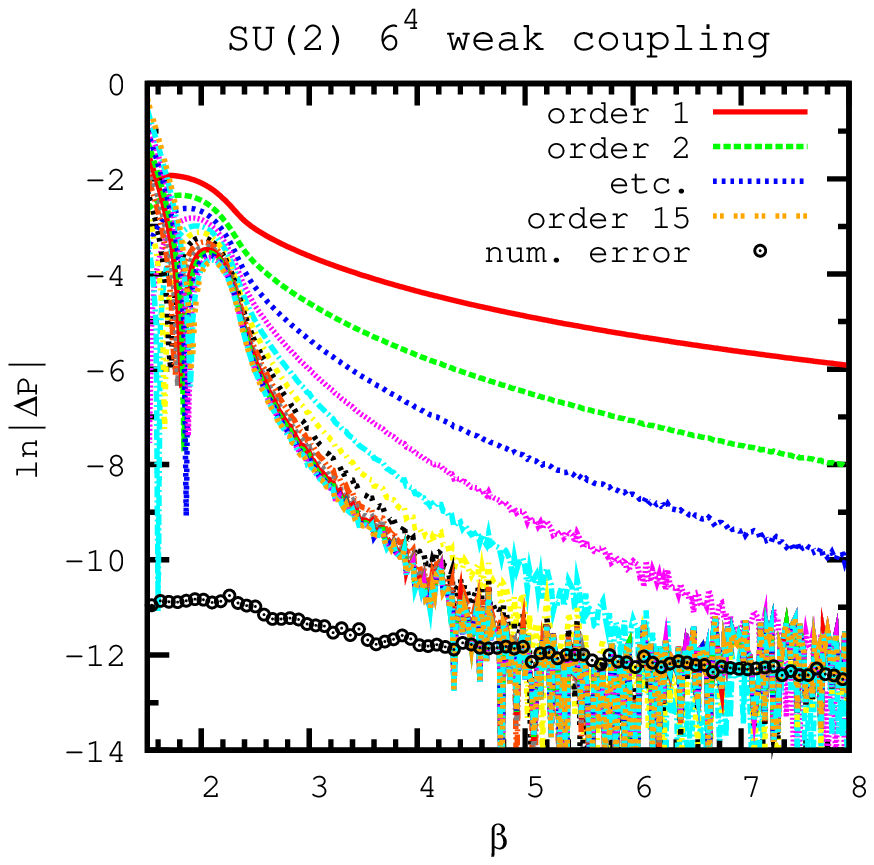}
\includegraphics[width=\columnwidth]{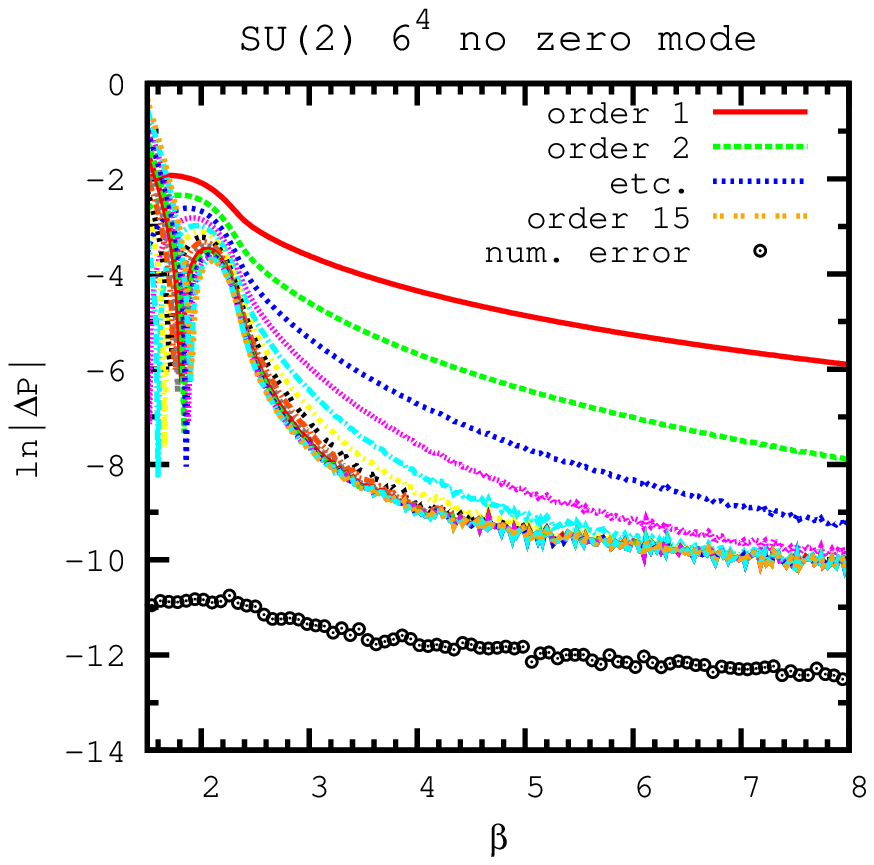}
\caption{
\label{fig:pldiff} Logarithm of the absolute value of the difference between the numerical data and the weak coupling expansion of $P$ at successive orders (above). For reference, we give the estimated numerical error on P. The graph below is the same except that we have not included the zero mode in $b_1$. }
\end{figure}

We have compared the weak coupling expansion of $P$ with numerical values in the case $V=6^4$. The results are shown in Fig. \ref{fig:pldiff}. In the region where the curves are smooth, the error decrease with the order and appears to accumulate. 
This is very similar to the case of $SU(3)$\cite{npp}.  However, it is clear that more 
reliable estimates for $m\geq 4$ would be desirable for $SU(2)$. It should be noted that 
for large $\beta$, the noise in the error is at the same level as the numerical error 
on $P$. This would not be the case if we had not included the contribution of the zero mode to $b_1$ as shown in the second part of Fig. \ref{fig:pldiff}.

\begin{figure}
\includegraphics[width=\columnwidth]{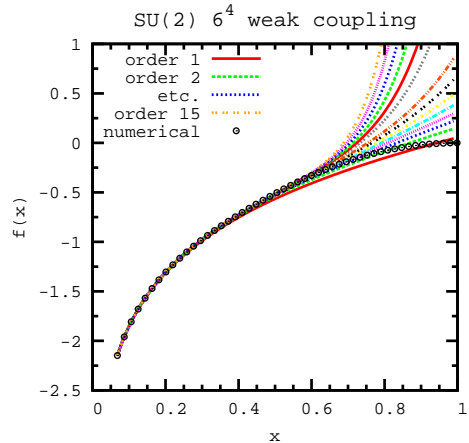}
\caption{
\label{fig:wd} Numerical value of $f(x)$ compared to the weak coupling expansion at 
successive orders.}
\end{figure}

\begin{figure}
\includegraphics[width=\columnwidth]{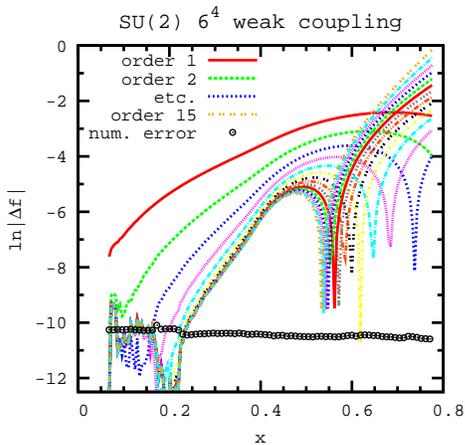}
\caption{
\label{fig:wddiff}Logarithm of the absolute value of the difference between the numerical data and the weak coupling expansion of $f$ at successive orders. For reference, we give the estimated numerical error on $f$.}
\end{figure}

We have compared the weak coupling expansion of $f(x)$ with numerical values in the case $V=6^4$. The results are shown in Fig. \ref{fig:wd}. The differences are resolved 
in Fig. \ref{fig:wddiff}. In these graphs we have taken $f_0=-0.14663$ which maximizes 
the length of the accumulation line on the left of Fig. \ref{fig:wddiff}.

\section{Expansion in Legendre polynomials}
\label{sec:leg}
\begin{figure}
\includegraphics[width=\columnwidth]{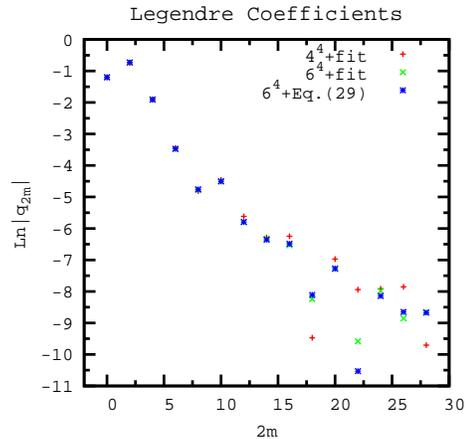}
\caption{
\label{fig:leg}  Legendre polynomial coefficients $q_{2m}$ with the three methods described in the text. }
\end{figure}

\begin{figure}
\includegraphics[width=\columnwidth]{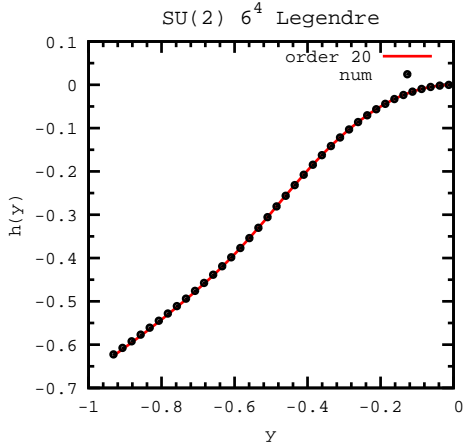}
\caption{
\label{fig:hx}  $h(y)$ together with the expansion in Legendre polynomials  up to order 20.  }
\end{figure}

We now consider the function $h(y)$, which is $g(y)$ with the logarithmic singularity 
subtracted as defined in Eq. (\ref{eq:hdef}).  This is a bell shaped even
function defined on the interval $[-1,1]$ and shown on Fig. \ref{fig:hx}. We can expand this function in terms of the 
even Legendre polynomials. 
\begin{equation}
h(y)=\sum_{m=0} q_{2m}P_{2m}(y)
\end{equation}
The $q_{2m}$ can be determined from the orthogonality relations with interpolated 
values of $h(y)$ to perform the integral. A minor technical difficulty is that we do not 
have numerical data all the way down to $y=-1$. This is 
because as $y\rightarrow  -1^{+}$, 
or in other words $x\rightarrow 0^{+}$,  $\beta\rightarrow +\infty$ where the plaquette distribution becomes infinitely narrow. Consequently there is a small gap 
in the numerical data that needs to be filled. Fortunately, this is precisely where the 
weak coupling expansion works well. Using the weak coupling expansion (including the 
overall constant), subtracting $Aln(x(2-x))$ and shifting to the $y$ coordinate, we 
obtained the approximate behavior near $y=-1$ for the $6^4$ data:
\begin{equation}
\label{eq:happ}
h(y)\simeq 0.2145+ 1.2961 y + 0.5261 y^2 + 0.1109 y^3
\end{equation}
In order to estimate the error associated with this approximation we have compared 
with an extrapolation of a quadratic fit of the leftmost part of the data. In order to 
give an idea of the volume effects, we have also used the second method on a $4^4$ 
lattice. The results are shown in Table \ref{tab:leg}. This indicates that the variations are 
small, increase with the order in relative magnitude and that the volume effects are 
stronger than the dependence on the extrapolation procedure. The logarithm of the coefficients is shown in Fig. \ref{fig:leg} which illustrate the exponential decay of the coefficients.
\begin{table}[b]
\begin{tabular}{||c|c|c|c||}
\hline
method& $4^4$ +fit&$6^4$ +fit &$6^4$+ (\ref{eq:happ})\cr
\hline
\hline
 $m$&$q_{2m}$&$q_{2m}$&$q_{2m}$\cr
 \hline
 0& -0.30034&  -0.30095&  -0.30096\cr  
1& -0.47963&  -0.48159&  -0.48164\cr  
2& 0.1488&  0.14853&  0.14845\cr  
3& -0.03215&  -0.0309&  -0.03099\cr  
4& -0.00822&  -0.00843&  -0.00852\cr  
5& 0.01156&  0.01114&  0.01107\cr  
6& -0.00363&  -0.00305&  -0.00308\cr  
7& -0.00186&  -0.00179&  -0.0018\cr  
8& 0.00194&  0.00146&  0.00147\cr  
9& 0.00008&  0.00026&  0.00028\cr  
10& -0.00094&  -0.00069&  -0.00067\cr   
\hline
\end{tabular}
\caption{\label{tab:leg} Legendre polynomial coefficients $q_{2m}$ with the three methods described in the text.}
\end{table}

\begin{figure}
\includegraphics[width=\columnwidth]{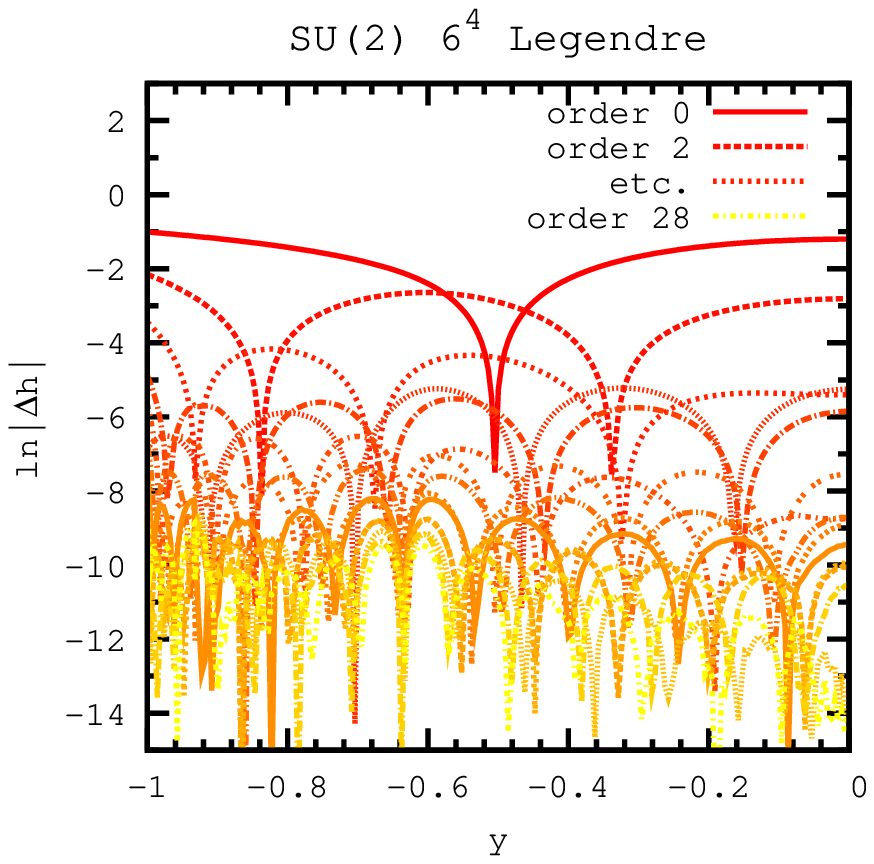}
\caption{
\label{fig:lnh}  Logarithm of the absolute value of the difference between the numerical data for $h(y)$ and expansions in Legendre polynomials at successive orders  }
\end{figure}

\begin{figure}
\includegraphics[width=\columnwidth]{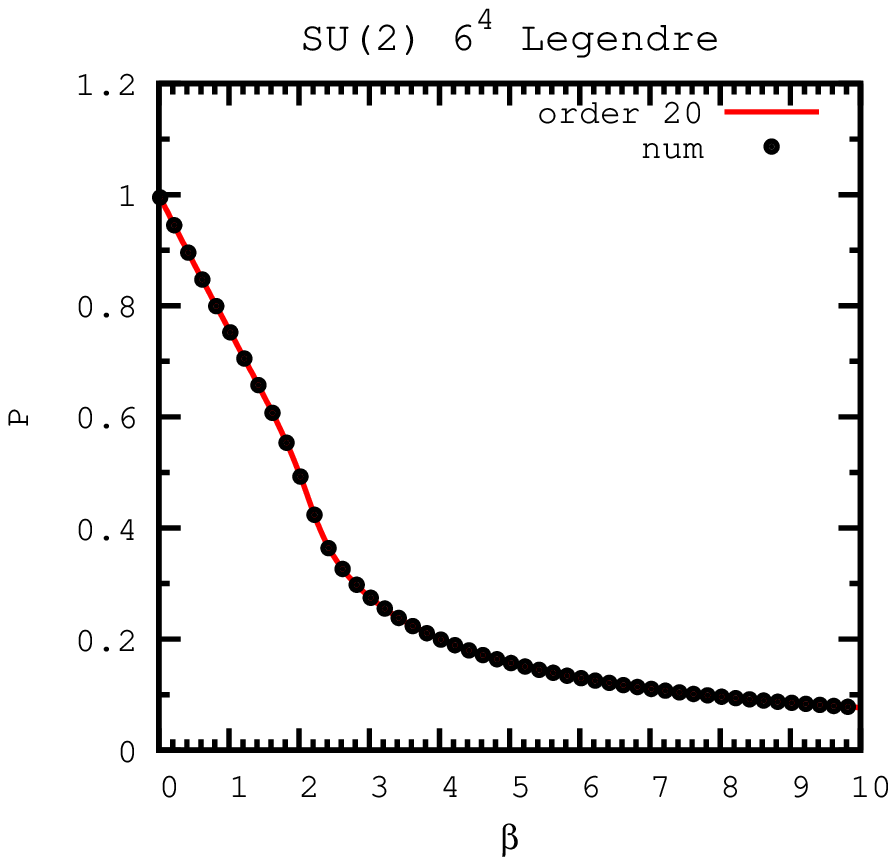}
\caption{
\label{fig:pb}    $P$ together with the expansion in Legendre polynomials  up to order 20.  }
\end{figure}

\begin{figure}
\includegraphics[width=\columnwidth]{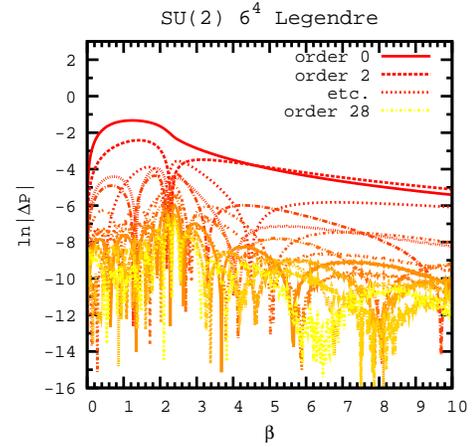}
\caption{
\label{fig:lnp}  Logarithm of the absolute value of the difference between the numerical data for $P$ and expansions in Legendre polynomials at successive orders  }
\end{figure}

The expansion provides excellent approximation of $h(y)$ shown in Fig. \ref{fig:hx}.
The errors are resolved in Fig. \ref{fig:lnh}. 
It is also possible to calculate $P(\beta)$ by solving the saddle point Eq. (\ref{eq:saddle})
using successive approximations for $h$. This is shown in Figs. \ref{fig:pb} and \ref{fig:lnp}. The spikes in the error graphs correspond to change of sign of the errors. 
It is important to notice that the quality of the approximations improves with the order in 
all region of the interval.

\section{Conclusions}
\begin{figure}
\includegraphics[width=\columnwidth]{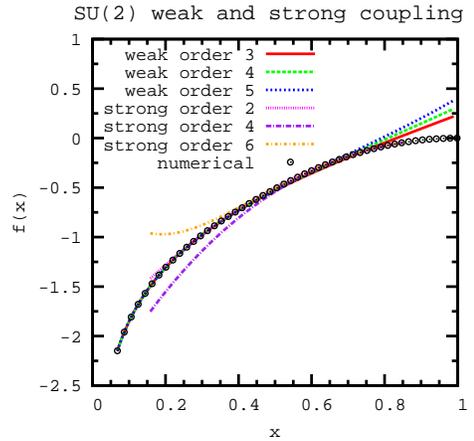}
\caption{
\label{fig:wall}  Weak and strong coupling expansion of $f$ at a few intermediate orders. }
\end{figure}

We have calculated the density of states for $SU(2)$ lattice gauge theory. 
The intermediate orders in weak and strong coupling agree well in an overlapping 
region of action values as shown in Fig. \ref{fig:wall}. However, the large order behaviors of these expansions 
appear to be similar to the corresponding ones for the plaquette. Volume effects 
can be resolved well for small actions values. Corrections to the saddle point estimate need to be developed systematically. Aprroximation of a subtracted quantity by Legendre 
polynomials looks very promising and works well uniformly. We plan to use 
this approximate form to look for Fisher's zeros. 

The density of states can be calculated 
in more general situations. For instance, 
\begin{equation}
Z(\beta,\{\beta_i\}) =\int_0^{2\mn}dS\ n(S,\{\beta_i\}){\rm e}^{-\beta S}\ ,
\label{eq:intdsg}
 \end{equation}
 with 
 \begin{eqnarray}
n(S,\{\beta_i\})&=&\\
\prod_{l}\int dU_l \delta(S&-&\sum_{p}(1-(1/N)Re Tr(U_p)))\\
&\times&{\rm e}^{-\sum_i\beta_i (1-\chi_i(U_p)/d_i)} \  ,
\end{eqnarray}
and $\chi_i$ a complete set of $SU(2)$ characters. This is a type of action which naturally arises in the RG studies of $SU(N)$ 
lattice gauge theories. 
It is possible to apply exact renormalization 
group transformation \cite{tomboulis07,tomboulis07b} or the MCRG procedure \cite{Tomboulis:2007rn}
to the partition function in order to define the couplings. 
Following the analogy between $f'=\beta$ 
and $V'=J$ for the effective potential $V$ in presence of a source $J$ in scalar models, 
it would be interesting to study finite size effects from this point of view.

\begin{acknowledgments}
This 
research was supported in part  by the Department of Energy
under Contract No. FG02-91ER40664. 
A.V. work was supported by the Joint Theory Institute funded together by 
Argonne National Laboratory and the University of Chicago, and in part by the U.S. Department of Energy, Division of High Energy Physics and Office of Nuclear Physics, under Contract DE-AC02-06CH11357.
\end{acknowledgments}
\bibliography{macbib}

\begin{thebibliography}{21}
\expandafter\ifx\csname natexlab\endcsname\relax\def\natexlab#1{#1}\fi
\expandafter\ifx\csname bibnamefont\endcsname\relax
  \def\bibnamefont#1{#1}\fi
\expandafter\ifx\csname bibfnamefont\endcsname\relax
  \def\bibfnamefont#1{#1}\fi
\expandafter\ifx\csname citenamefont\endcsname\relax
  \def\citenamefont#1{#1}\fi
\expandafter\ifx\csname url\endcsname\relax
  \def\url#1{\texttt{#1}}\fi
\expandafter\ifx\csname urlprefix\endcsname\relax\def\urlprefix{URL }\fi
\providecommand{\bibinfo}[2]{#2}
\providecommand{\eprint}[2][]{\url{#2}}

\bibitem[{\citenamefont{Tomboulis}(2007{\natexlab{a}})}]{tomboulis07}
\bibinfo{author}{\bibfnamefont{E.~T.} \bibnamefont{Tomboulis}}
  (\bibinfo{year}{2007}{\natexlab{a}}), \eprint{0707.2179}.

\bibitem[{\citenamefont{Tomboulis}(2007{\natexlab{b}})}]{tomboulis07b}
\bibinfo{author}{\bibfnamefont{E.~T.} \bibnamefont{Tomboulis}},
  \bibinfo{journal}{PoS} \textbf{\bibinfo{volume}{LATTICE2007}},
  \bibinfo{pages}{336} (\bibinfo{year}{2007}{\natexlab{b}}),
  \eprint{0710.1894}.

\bibitem[{\citenamefont{Kogut}(1980)}]{kogut80b}
\bibinfo{author}{\bibfnamefont{J.~B.} \bibnamefont{Kogut}},
  \bibinfo{journal}{Phys. Rept.} \textbf{\bibinfo{volume}{67}},
  \bibinfo{pages}{67} (\bibinfo{year}{1980}).

\bibitem[{\citenamefont{Li and Meurice}(2006)}]{third}
\bibinfo{author}{\bibfnamefont{L.}~\bibnamefont{Li}} \bibnamefont{and}
  \bibinfo{author}{\bibfnamefont{Y.}~\bibnamefont{Meurice}},
  \bibinfo{journal}{Phys. Rev.} \textbf{\bibinfo{volume}{D73}},
  \bibinfo{pages}{036006} (\bibinfo{year}{2006}), \eprint{hep-lat/0507034}.

\bibitem[{\citenamefont{Li and Meurice}(2005{\natexlab{a}})}]{plaquette}
\bibinfo{author}{\bibfnamefont{L.}~\bibnamefont{Li}} \bibnamefont{and}
  \bibinfo{author}{\bibfnamefont{Y.}~\bibnamefont{Meurice}},
  \bibinfo{journal}{Phys. Rev.} \textbf{\bibinfo{volume}{D71}},
  \bibinfo{pages}{054509} (\bibinfo{year}{2005}{\natexlab{a}}),
  \eprint{hep-lat/0501023}.

\bibitem[{\citenamefont{Alves et~al.}(1990{\natexlab{a}})\citenamefont{Alves,
  Berg, and Sanielevici}}]{alves90b}
\bibinfo{author}{\bibfnamefont{N.~A.} \bibnamefont{Alves}},
  \bibinfo{author}{\bibfnamefont{B.~A.} \bibnamefont{Berg}}, \bibnamefont{and}
  \bibinfo{author}{\bibfnamefont{S.}~\bibnamefont{Sanielevici}},
  \bibinfo{journal}{Phys. Rev. Lett.} \textbf{\bibinfo{volume}{64}},
  \bibinfo{pages}{3107} (\bibinfo{year}{1990}{\natexlab{a}}).

\bibitem[{\citenamefont{Denbleyker
  et~al.}(2007{\natexlab{a}})\citenamefont{Denbleyker, Du, Meurice, and
  Velytsky}}]{quasig}
\bibinfo{author}{\bibfnamefont{A.}~\bibnamefont{Denbleyker}},
  \bibinfo{author}{\bibfnamefont{D.}~\bibnamefont{Du}},
  \bibinfo{author}{\bibfnamefont{Y.}~\bibnamefont{Meurice}}, \bibnamefont{and}
  \bibinfo{author}{\bibfnamefont{A.}~\bibnamefont{Velytsky}},
  \bibinfo{journal}{Phys. Rev.} \textbf{\bibinfo{volume}{D76}},
  \bibinfo{pages}{116002} (\bibinfo{year}{2007}{\natexlab{a}}),
  \eprint{arXiv:0708.0438 [hep-lat]}.

\bibitem[{\citenamefont{Denbleyker
  et~al.}(2007{\natexlab{b}})\citenamefont{Denbleyker, Du, Meurice, and
  Velytsky}}]{lat07}
\bibinfo{author}{\bibfnamefont{A.}~\bibnamefont{Denbleyker}},
  \bibinfo{author}{\bibfnamefont{D.}~\bibnamefont{Du}},
  \bibinfo{author}{\bibfnamefont{Y.}~\bibnamefont{Meurice}}, \bibnamefont{and}
  \bibinfo{author}{\bibfnamefont{A.}~\bibnamefont{Velytsky}},
  \bibinfo{journal}{PoS} \textbf{\bibinfo{volume}{LAT2007}},
  \bibinfo{pages}{269} (\bibinfo{year}{2007}{\natexlab{b}}),
  \eprint{0710.5771}.

\bibitem[{\citenamefont{Li and Meurice}(2005{\natexlab{b}})}]{gluodyn04}
\bibinfo{author}{\bibfnamefont{L.}~\bibnamefont{Li}} \bibnamefont{and}
  \bibinfo{author}{\bibfnamefont{Y.}~\bibnamefont{Meurice}},
  \bibinfo{journal}{Phys. Rev. D} \textbf{\bibinfo{volume}{71}},
  \bibinfo{pages}{016008} (\bibinfo{year}{2005}{\natexlab{b}}),
  \eprint{hep-lat/0410029}.

\bibitem[{\citenamefont{Alves et~al.}(1990{\natexlab{b}})\citenamefont{Alves,
  Berg, and Villanova}}]{alves89}
\bibinfo{author}{\bibfnamefont{N.~A.} \bibnamefont{Alves}},
  \bibinfo{author}{\bibfnamefont{B.~A.} \bibnamefont{Berg}}, \bibnamefont{and}
  \bibinfo{author}{\bibfnamefont{R.}~\bibnamefont{Villanova}},
  \bibinfo{journal}{Phys. Rev.} \textbf{\bibinfo{volume}{B41}},
  \bibinfo{pages}{383} (\bibinfo{year}{1990}{\natexlab{b}}).

\bibitem[{\citenamefont{Alves et~al.}(1992)\citenamefont{Alves, Berg, and
  Sanielevici}}]{alves91}
\bibinfo{author}{\bibfnamefont{N.~A.} \bibnamefont{Alves}},
  \bibinfo{author}{\bibfnamefont{B.~A.} \bibnamefont{Berg}}, \bibnamefont{and}
  \bibinfo{author}{\bibfnamefont{S.}~\bibnamefont{Sanielevici}},
  \bibinfo{journal}{Nucl. Phys.} \textbf{\bibinfo{volume}{B376}},
  \bibinfo{pages}{218} (\bibinfo{year}{1992}), \eprint{hep-lat/9107002}.

\bibitem[{\citenamefont{Balian et~al.}(1975)\citenamefont{Balian, Drouffe, and
  Itzykson}}]{balian74}
\bibinfo{author}{\bibfnamefont{R.}~\bibnamefont{Balian}},
  \bibinfo{author}{\bibfnamefont{J.~M.} \bibnamefont{Drouffe}},
  \bibnamefont{and} \bibinfo{author}{\bibfnamefont{C.}~\bibnamefont{Itzykson}},
  \bibinfo{journal}{Phys. Rev.} \textbf{\bibinfo{volume}{D11}},
  \bibinfo{pages}{2104} (\bibinfo{year}{1975}).

\bibitem[{\citenamefont{Balian et~al.}(1979)\citenamefont{Balian, Drouffe, and
  Itzykson}}]{balian74err}
\bibinfo{author}{\bibfnamefont{R.}~\bibnamefont{Balian}},
  \bibinfo{author}{\bibfnamefont{J.~M.} \bibnamefont{Drouffe}},
  \bibnamefont{and} \bibinfo{author}{\bibfnamefont{C.}~\bibnamefont{Itzykson}},
  \bibinfo{journal}{Phys. Rev.} \textbf{\bibinfo{volume}{D19}},
  \bibinfo{pages}{2514} (\bibinfo{year}{1979}).

\bibitem[{\citenamefont{Li and Meurice}(2004)}]{anlqcd}
\bibinfo{author}{\bibfnamefont{L.}~\bibnamefont{Li}} \bibnamefont{and}
  \bibinfo{author}{\bibfnamefont{Y.}~\bibnamefont{Meurice}}
  (\bibinfo{year}{2004}), \eprint{hep-lat/0411020}.

\bibitem[{\citenamefont{Heller and Karsch}(1985)}]{heller84}
\bibinfo{author}{\bibfnamefont{U.~M.} \bibnamefont{Heller}} \bibnamefont{and}
  \bibinfo{author}{\bibfnamefont{F.}~\bibnamefont{Karsch}},
  \bibinfo{journal}{Nucl. Phys.} \textbf{\bibinfo{volume}{B251}},
  \bibinfo{pages}{254} (\bibinfo{year}{1985}).

\bibitem[{\citenamefont{Coste et~al.}(1985)\citenamefont{Coste,
  Gonzalez-Arroyo, Jurkiewicz, and Korthals~Altes}}]{coste85}
\bibinfo{author}{\bibfnamefont{A.}~\bibnamefont{Coste}},
  \bibinfo{author}{\bibfnamefont{A.}~\bibnamefont{Gonzalez-Arroyo}},
  \bibinfo{author}{\bibfnamefont{J.}~\bibnamefont{Jurkiewicz}},
  \bibnamefont{and} \bibinfo{author}{\bibfnamefont{C.~P.}
  \bibnamefont{Korthals~Altes}}, \bibinfo{journal}{Nucl. Phys.}
  \textbf{\bibinfo{volume}{B262}}, \bibinfo{pages}{67} (\bibinfo{year}{1985}).

\bibitem[{\citenamefont{Alles et~al.}(1994)\citenamefont{Alles, Campostrini,
  Feo, and Panagopoulos}}]{alles93}
\bibinfo{author}{\bibfnamefont{B.}~\bibnamefont{Alles}},
  \bibinfo{author}{\bibfnamefont{M.}~\bibnamefont{Campostrini}},
  \bibinfo{author}{\bibfnamefont{A.}~\bibnamefont{Feo}}, \bibnamefont{and}
  \bibinfo{author}{\bibfnamefont{H.}~\bibnamefont{Panagopoulos}},
  \bibinfo{journal}{Phys. Lett.} \textbf{\bibinfo{volume}{B324}},
  \bibinfo{pages}{433} (\bibinfo{year}{1994}), \eprint{hep-lat/9306001}.

\bibitem[{\citenamefont{Di~Renzo and Scorzato}(2001)}]{direnzo2000}
\bibinfo{author}{\bibfnamefont{F.}~\bibnamefont{Di~Renzo}} \bibnamefont{and}
  \bibinfo{author}{\bibfnamefont{L.}~\bibnamefont{Scorzato}},
  \bibinfo{journal}{JHEP} \textbf{\bibinfo{volume}{10}}, \bibinfo{pages}{038}
  (\bibinfo{year}{2001}), \eprint{hep-lat/0011067}.

\bibitem[{\citenamefont{Rakow}(2006)}]{rakow05}
\bibinfo{author}{\bibfnamefont{P.~E.~L.} \bibnamefont{Rakow}},
  \bibinfo{journal}{PoS} \textbf{\bibinfo{volume}{LAT2005}},
  \bibinfo{pages}{284} (\bibinfo{year}{2006}), \eprint{hep-lat/0510046}.

\bibitem[{\citenamefont{Meurice}(2006)}]{npp}
\bibinfo{author}{\bibfnamefont{Y.}~\bibnamefont{Meurice}},
  \bibinfo{journal}{Phys. Rev.} \textbf{\bibinfo{volume}{D74}},
  \bibinfo{pages}{096005} (\bibinfo{year}{2006}), \eprint{hep-lat/0609005}.

\bibitem[{\citenamefont{Tomboulis and Velytsky}(2007)}]{Tomboulis:2007rn}
\bibinfo{author}{\bibfnamefont{E.~T.} \bibnamefont{Tomboulis}}
  \bibnamefont{and} \bibinfo{author}{\bibfnamefont{A.}~\bibnamefont{Velytsky}},
  \bibinfo{journal}{Phys. Rev.} \textbf{\bibinfo{volume}{D75}},
  \bibinfo{pages}{076002} (\bibinfo{year}{2007}), \eprint{hep-lat/0702015}.

\end{thebibliography}

\end{document}